\documentclass[twocolumn,pra,showpacs,amsmath,superscriptaddress]{revtex4-2}
\usepackage{epsfig,graphicx}
\usepackage{bookmark}
\usepackage{braket}
\usepackage{xcolor}%{\color{red} text}

\newcommand{\be}{\begin{equation}}
\newcommand{\e}[1]{\label{#1}\end{equation}}
\def\bea{\begin{eqnarray}}
\def\ea#1{\label{#1}\end{eqnarray}}
\def\rqn#1{(\ref{#1})}
\def\ee{\end{equation}}
\def\eea{\end{eqnarray}}
\def\bes#1{\begin{subequations}\label{#1}}
\def\ese{\end{subequations}}

\begin{document}

\title{Quantum Zeno and Anti-Zeno probes of noise correlations in photon polarisation}

\author{Salvatore Virz\`i}
\author{Alessio Avella}
\author{Fabrizio Piacentini}
\author{Marco Gramegna}
\affiliation{INRIM - Istituto Nazionale di Ricerca Metrologica, strada delle Cacce, 91 10135 Torino, Italy.}

\author{Tom\'{a}\v{s} Opatrn\'{y}}
\affiliation{Department of Optics, Palack\'{y} University, 77200 Olomouc, Czech Republic.}

\author{Abraham G. Kofman}
\affiliation{Department of Chemical and Biological Physics, Weizmann Institute of Science, Rehovot 7610001, Israel.}
\affiliation{Department of  Physics, Shanghai University, 200444 Shanghai, China.}

\author{Gershon Kurizki}
\affiliation{Department of Chemical and Biological Physics, Weizmann Institute of Science, Rehovot 7610001, Israel.}

\author{Stefano Gherardini}
\author{Filippo Caruso}
\affiliation{Dept. of Physics and Astronomy and European Laboratory for Non-Linear Spectroscopy (LENS), University of Florence, via G. Sansone 1, 50019 Sesto Fiorentino, Italy.}

\author{Ivo Pietro Degiovanni}
\author{Marco Genovese}
\affiliation{INRIM - Istituto Nazionale di Ricerca Metrologica, strada delle Cacce, 91 10135 Torino, Italy.}
\affiliation{INFN, sede di Torino, via P. Giuria 1, 10125 Torino, Italy.}

\date{\today}

\begin{abstract}
We experimentally demonstrate, for the first time, noise diagnostics by repeated quantum measurements.
Specifically, we establish  the ability of a single photon, subjected to  random polarisation noise, to diagnose non-Markovian temporal correlations of such a  noise process.
In the frequency domain, these noise correlations correspond to colored noise spectra, as opposed to the ones related to Markovian, white noise.
Both the noise spectrum and its corresponding temporal correlations  are diagnosed by probing the photon by means of frequent, (partially-)selective polarisation measurements.
Our main result is the experimental demonstration that noise with positive temporal correlations corresponds to our single photon undergoing a dynamical regime enabled by the quantum Zeno effect (QZE), while noise characterized by negative (anti-) correlations corresponds to regimes associated with the anti-Zeno effect (AZE).
This demonstration opens the way to a new kind of noise spectroscopy based on QZE and AZE in photon (or other single-particle) state probing.
\end{abstract}

%\pacs{03.65.Ta, 03.65.Ca, 03.65.Wj}
\maketitle

%\noindent
%From: \\
%To: \\

%\tableofcontents

\section{Introduction}

Quantum control \cite{qco1,qco2,qco3} is a fundamental tool for the development of highly accurate quantum technologies.
In particular, the quantum Zeno (QZE) \cite{1,2,3,4} and anti-Zeno (AZE) \cite{5} effects, respectively denoting the slowdown and speedup of quantum-system evolution by its frequent interruptions \cite{6,7,8,9,10,11,13,14,15}, %\cite{1,2,3,4,5,6,7,8,9,10,11,12,13,14,15,16,17,18,19,20,21,22,22a,23},
have been recognised (beyond their fundamental significance) to represent general paradigms of quantum control \cite{16,17,18,19,20,22,23}.
Indeed, they allow to either protect \cite{12} or steer the quantum state of a system via an interplay between the effects of frequent operations (system control) and the coupling of the system to its environment (a bath) \cite{21,22,22a,23,23a,23b}.

The generality of these paradigms is revealed by the Kofman-Kurizki (KK) universal formula, whereby the overlap of the system-bath coupling spectrum with the system-control spectrum determines the decay (relaxation) rate $\gamma(t)$ of the initial-state population \cite{5,22,22a,23,23a,23b,23c}
\be
\gamma(t) =2\pi\int_{-\infty}^\infty d\omega\,G(\omega)F_t(\omega).
\e{1}
In Eq. \rqn{1} $G(\omega)$ is the spectrum of the system-bath coupling (bath-response) and $F_t(\omega)$  the system control spectrum evaluated within the time interval $[0,t]$.
According to Eq. \rqn{1}, the QZE  corresponds to the suppression of the bath-induced decay $\gamma(t)$ by the reduction of the overlap between the chosen $F_t(\omega)$ and the bath response $G(\omega)$, whereas the AZE corresponds to the enhancement of $\gamma(t)$ by increasing this overlap \cite{5,22,23}.
This means that the time-variation of  the system control must be much faster than (in the QZE case) or as fast as (in the AZE case) the bath correlation time, with the result that both effects are distinctly non-Markovian.
Overall, the only condition on the validity of Eq. \rqn{1} is the weakness of the system-bath coupling, allowing for a perturbative treatment of the bath effects.

The KK formula has been confirmed, theoretically and experimentally, in scenarios involving frequent perturbations of open-system evolution, such as the dynamics of cold atoms in optical lattices \cite{6,22}, the trapping of Bose-Einstein condensates \cite{q1,q2,q3}, light propagation in waveguides \cite{q4a,q4b} and cavities \cite{4,q4a,q4b},  AZE-cooling and  QZE-heating of qubits coupled to a bath \cite{q5,q6,q7,q8}.
Moreover, Eq. \rqn{1} can be  used  for designing  optimal  protection of  multi-qubit quantum information processing \cite{q9} or transfer mechanisms \cite{q10}.

In this paper, we study both theoretically and experimentally an alternative purpose of the KK formula, i.e. the diagnostics (characterization)  of  random processes, alias noise spectroscopy \cite{q11}.
One may infer the bath-response spectrum $G(\omega)$ upon varying the control spectrum $F_t(\omega)$ and recording the resulting decoherence rate, as it has already been experimentally confirmed in \cite{q12,q13}.
This, however, is a time-consuming process.
Alternatively, key information on the bath-response spectrum, such as its width (the inverse memory time), may be gathered by appropriate dynamical control of the system that probes the bath \cite{q11}.
This represents an innovative and powerful tool that we introduce to expand quantum sensing technology.
Although, in this context, the QZE has been previously used both experimentally and theoretically \cite{39,40} as a means of assisting noise sensing, here we demonstrate its ability to serve as a source of information on noise processes, a direction that has not been investigated as yet and is of high relevance to quantum technologies.

More specifically, we experimentally demonstrate, for the first time, noise diagnostics by repeated quantum measurements.
In particular, we establish the ability of a single photon, undergoing random polarisation fluctuations, to diagnose non-Markovian temporal correlations within this noise.
In the frequency domain, Markovian and non-Markovian noise correlations correspond, respectively, to white and colored noise spectra.
We  show that the noise spectrum, as well as its underlying temporal correlations, is diagnosed when the photon is frequently probed by polarisation measurements.
The main purpose of this work is the experimental demonstration that a noise characterized by positive (negative) time-correlations gives rise to QZE (AZE). This result  paves the way to a new generation of noise spectroscopy protocols based on QZE and AZE in photon (or other single-particle) state probing.

\section{Theoretical Model}

%After its initialization in the horizontally polarized state $\ket{H}$,
Let us consider a single photon that, initialized in the horizontally-polarised state $\ket{H}$, passes through a sequence of $N$ blocks at time instants $t_1,\dots,t_N$.

\begin{figure}[htb]
\includegraphics[width=7cm]{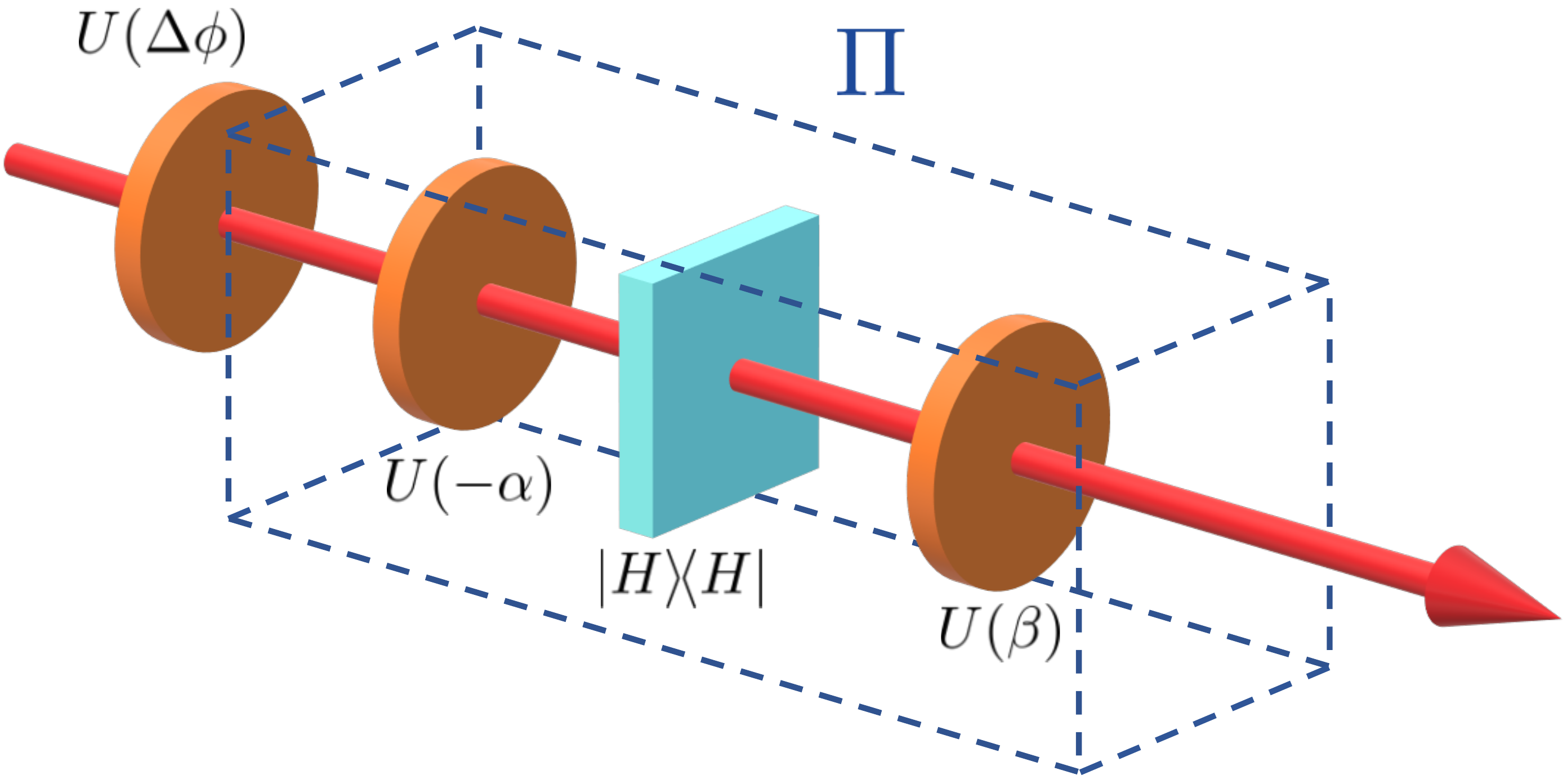}
%\centerline{\epsfig{file=garda7.ps,width=3in}}
\protect\caption{Scheme of the $k$-th element in a sequence of $N$ blocks determining the quantum state evolution of the photon passing through them.
First, the polarisation state of the photon undergoes a random rotation $U(\Delta\phi_k)$. Then, the unitary rotation $U(-\alpha_k)$, followed by the projection onto the state $\ket{H}\bra{H}$ and by a proper counter-rotation $U (\beta_k)$, realises the selective measurement $\Pi$ defined in Eq. (\ref{2}).}
\label{fig1}\end{figure}

In the $k$-th block (Fig. \ref{fig1}, $k=1,...,N$), the state of the photon is rotated by the operator $U(\Delta\phi_k)$, representing a random  polarisation rotation in the $x$-$z$ plane of the Bloch sphere (i.e. a $\hat{\sigma}_y$ Pauli-matrix rotation around the $y$-axis):
%\bea
%&U(\Delta\phi_k)&=\exp(-i\hat{\sigma}_y\Delta\phi_k)
%\nonumber\\
%&&=I\cos\Delta\phi_k-i\hat{\sigma}_y\sin\Delta\phi_k,
%\ea{6}
\be
U(\Delta\phi_k)=e^{-i\hat{\sigma}_y\Delta\phi_k}=I\cos\Delta\phi_k-i\hat{\sigma}_y\sin\Delta\phi_k,
\e{6}
Here $I$ denotes the two-dimensional identity matrix, $U(\Delta\phi)$ describes a rotation by $2\Delta\phi$ on the Bloch sphere, which corresponds to a rotation by the angle $\Delta\phi$ of the photon linear polarisation.

By passing through a sequence of $N$ distinct blocks, the horizontal ($\ket{H}$) and vertical ($\ket{V}$) polarisation states of the photon evolve as the degenerate states of a two-level system coupled by intermittent polarisation rotations.
These rotations are interspersed by an equivalent number of selective measurements.
More specifically, in each of the $N$ blocks the photon state undergoes a quantum measurement that corresponds to partial or complete absorption of the vertical polarisation component (or, equivalently, to partial projection on the horizontal polarisation state), i.e.:
\be
\Pi=\ket{H}\bra{H}+\theta\ket{V}\bra{V} =
\theta I + (1-\theta)\ket{H}\bra{H},
\e{2}
where $\theta\in[0,1]$ is the parameter determining the measurement selective strength.
The quantum measurement in Eq. \rqn{2} can be realised exploiting the polarisation rotation $U(-\alpha_k)$, followed by the projection $\ket{H}\bra{H}$ and the second polarisation rotation $U(\beta_k)$.
More specifically, $U(-\alpha_k)$ determines the selective strength of the measurement $\Pi$, while $U(\beta_k)$ fixes the polarisation of the photon exiting the $k$-th block.\\
%Such measurement effects are reproduced by means of the rotation $U(-\alpha_k)$, followed by the projection $\ket{H}\bra{H}$ and the second rotation $U(\beta_k)$.
%Here, the presence of $U(-\alpha_k)$ models the losses caused by the selective measurement $\Pi$, while $U(\beta_k)$ fixes the polarisation of the photon exiting the $k$-th block.\\
%At the exit from the $k$-th block, one has the following (unnormalized) state:
%\be
%\ket{\psi(t_k)}=\left(\prod_{l=1}^k[\Pi U(\Delta\phi_l)]\right)\ket{H},
%\e{3}
%with the corresponding survival probability of the horizontally-polarised photon being
%\be
%P_H(t_k) =|\braket{H|\psi(t_k)}|^2.
%\e{4}
The output of the $k$-th block is the (unnormalized) state $\ket{\psi(t_k)}=\left(\prod_{l=1}^k[\Pi U(\Delta\phi_l)]\right)\ket{H}$.
Since the sequence $\Delta\phi_1,\dots,\Delta\phi_N$ is random, because of the external noise process, we are interested in the survival probability of the horizontal polarisation, averaged over all possible random sequences:
\be
\bar{P}_H(t_k) =\overline{|\braket{H|\psi(t_k)}|^2},
\e{7}
where the averaging operation is denoted by the overline.

Let us consider the case in which positive (or negative) polarisation rotation jumps with identical magnitude occur in the block sequence, i.e. $\left|\Delta\phi_k\right|=\Delta\phi$ $\forall k=1,...,N$.
%We can introduce a correlation parameter ${\cal C}$, describing the degree of correlation between consecutive phase jumps, defined as
%\be
%\overline{\Delta\phi}_{\Delta\phi'}={\cal C}\Delta\phi',
%\ee
%where $\overline{\Delta\phi}_{\Delta\phi'}$ is the average phase jump provided the previous jump was $\Delta\phi'$.
%We may write
%\be
%\overline{\Delta\phi}_{\Delta\phi'}=p\Delta\phi'-(1-p)\Delta\phi'
%=(2p-1)\Delta\phi',
%\ee
%which yields the relation
%\be
%{\cal C}=2p-1.
%\e{5}
Then, we can introduce the correlation parameter ${\cal C}$, describing the degree of correlation between consecutive polarisation rotation jumps.
It is defined as
\be
\overline{\Delta\phi}_{\Delta\phi'}=p\Delta\phi'-(1-p)\Delta\phi'={\cal C}\Delta\phi',
\ee
where $\overline{\Delta\phi}_{\Delta\phi'}$ is the average polarisation rotation jump at the end of the $k$-th block (provided the previous jump was $\Delta\phi'$) and $p$ denotes the correlation probability between subsequent jumps.
Thus, ${\cal C}=2p-1$, such that the jumps are correlated for ${\cal C}>0$ ($p>0.5$), anti-correlated for ${\cal C}<0$ ($p<0.5$), and uncorrelated for ${\cal C}=0$ ($p=0.5$).
A universal closed-form solution for the average survival probability in Eq. \rqn{7} is not available.
However, in the following subsections an analytical description of all the cases of interest will be given.
%A closed-form solution for the average survival probability in Eq. \rqn{7}, valid for all values of the problem parameters, is not available.
%However, it is possible to describe analytically all the cases of interest, as follows.

\subsection{Non-random evolution}
%\textit{Non-random evolution.}
In the maximally-correlated case $\mathcal{C}=1$, the time dependence of the survival probability $\bar{P}_H(t_k)$ is the same as for the non-random evolution with identical polarisation rotation jumps $\Delta\phi$ or ($-\Delta\phi$) in each block.
In particular, as proven in \cite{21}, the probability of observing the photon with a horizontal polarisation assumes the following closed form for any $\theta$:
\be
\bar{P}_H(t_k)=\frac{[\lambda_+^k(\cos\Delta\phi-\lambda_-)+
\lambda_-^k(\lambda_+-\cos\Delta\phi)]^2}
{(1+\theta)^2\cos^2\Delta\phi-4\theta},
\e{11}
where $\lambda_{\pm}\equiv\frac12\left[(1+\theta)\cos\Delta\phi
\pm\sqrt{(1+\theta)^2\cos^2\Delta\phi-4\theta}\right]$.
%\be
%P_H(t_k)=\frac{[\lambda_+^\frac{t_k}{\tau}(\cos\Delta\phi-\lambda_-)+
%\lambda_-^\frac{t_k}{\tau}(\lambda_+-\cos\Delta\phi)]^2}
%{(1+\theta)^2\cos^2\Delta\phi-4\theta},
%\e{11}
%with $\lambda_{\pm}=\frac12\left[(1+\theta)\cos\Delta\phi
%\pm\sqrt{(1+\theta)^2\cos^2\Delta\phi-4\theta}\right]$ and assuming that the blocks are equidistant, so that $t_k=k\tau$, where $\tau$ is the photon flight time between two consecutive blocks.
In the absence of measurements ($\theta=1$), one gets $\bar{P}_H(t_k)=\cos^{2}(k\Delta\phi)$, which is analogous to Rabi oscillations of the horizontal polarisation probability as a function of the number of blocks ($k$) traversed by the photon.
The opposite limit $\theta=0$ corresponds to the projective measurement case.
These measurements are {\em selective}, since each time the single photon passes through one of the blocks depicted in Fig. \ref{fig1} a projection of its polarisation state onto the horizontal polarisation occurs.
In this case, $\bar{P}_H(t_k)=\cos^{2k}(\Delta\phi)$, which is insensitive to sign fluctuations of the polarisation rotation jumps and hence is the same for random and non-random evolution.
Thus, it takes the form $\bar{P}_H(t_k)=e^{-k(\Delta\phi)^2}$ for small rotation angles ($\Delta\phi\ll1$).
This decay is slower than the period of uninterrupted Rabi oscillations, thus implying that QZE occurs.

Next, let us consider non-random evolution in the intermediate case $0<\theta<1$, corresponding to partially-selective measurements \cite{wrev}.
For small rotation angles and sufficiently selective measurements ($\Delta\phi\ll1-\theta$), Eq.~\rqn{11} reduces to the exponential decay $\bar{P}_H(t_k)=\exp\left[-\frac{(\Delta\phi)^2}{\tau^2\nu}t_k\right]$, where $\nu\equiv\frac{1-\theta}{1+\theta}\frac{1}{\tau}$ and the $N$ blocks are assumed to be equidistant (i.e. $t_k=k\tau$), with $\tau$ denoting the photon flight time between two consecutive blocks.

Note that the quantity $\nu$ is the (effective) measurement rate, i.e., the reciprocal time during which partially-selective measurements perform state selection.
Obviously, $\nu$ scales as $1/\tau$, and decreases as the selective strength parameter $\theta$ increases.
The decay rate {\em diminishes} with $\nu$, highlighting the QZE.

\subsection{Random evolution}
%\textit{Random evolution.}

%Now, let us take into account the case of random $(-1\le \mathcal{C}<1)$ rotation angles caused by noisy modulation of the polarisation rotation $U(\Delta\phi)$.
Now, let us consider the case of random noisy modulation of the polarisation rotation $U(\Delta\phi)$, both for correlated and anticorrelated noise $(-1\le \mathcal{C}<1)$.
%In the absence of measurements ($\theta=1$), the average uninterrupted polarisation evolution can be shown \cite{21} to be
%\be
%\bar{P}_H(t_k) =\frac12+\frac{g(r)-g(-r)}{4r},
%\e{8}
%where $g(r)=[r+(1-p)\cos(2\Delta\phi)][r+p\cos(2\Delta\phi)]^k$ and $r=\sqrt{(1-p)^2-p^2\sin^2(2\Delta\phi)}$.
In the general case of $0\le\theta\le1$, for small rotation angles $\Delta\phi\ll1$ (corresponding to our case of interest) one finds the solution \cite{21}
\be
\bar{P}_H(t_k)=e^{-(\gamma+\Gamma_0)t_k}\left(\cosh(St_k)+
\frac{\Gamma_0}{S}\sinh(St_k)\right),
\e{8.65'}
where $S\equiv\sqrt{\gamma^2+\Gamma_0^2}$, with $\gamma\equiv\frac{1+{\cal C}\theta}{1-{\cal {\cal C}}\theta}
\frac{\Delta\phi^2}{\tau}$ being the polarisation decay rate and $\Gamma_0\equiv-\frac{\ln\theta}{2\tau}$ denoting the time-averaged rate of photons absorbed by the polarisers.\\
In the absence of selective measurement ($\theta=1$), Eq. \rqn{8.65'} yields
%\be
%\bar{P}_H(t_k)=\frac{1+e^{-2\gamma_0t_k}}{2} \;\;\;\;\; \left(\gamma_0=\frac{1+{\cal C}}{1-{\cal C}}\frac{\Delta\phi^2}{\tau}\right).
%\e{8.65}
\be
\bar{P}_H(t_k)=\frac{1+e^{-2\gamma_0t_k}}{2},
\e{8.65}
where $\gamma_0\equiv\gamma|_{\theta=1}$.
Thus, random fluctuations of the polarisation lead to a completely unpolarised state, given by the limit $\bar{P}_H(t)\rightarrow1/2$ for $t\rightarrow\infty$.\\
For weakly-selective measurements ($\theta\rightarrow1$), in the limit $\Gamma_0\ll\gamma$ Eq. \rqn{8.65'} takes the form
\be
\bar{P}_H(t_k)\approx e^{-\gamma t_k},
\e{8.67}
showing an exponential decay over time of the horizontal polarisation probability.\\
Finally, in the scenario involving projective measurements ($\theta\to0$), the evolution of $\bar{P}_H(t_k)$ follows Eq. \rqn{8.67}, with $\gamma\rightarrow\Delta\phi^2/\tau$.\\

\section{Experiment}
%\textit{Experiment.}

\begin{figure}[htb]
\includegraphics[width=\columnwidth]{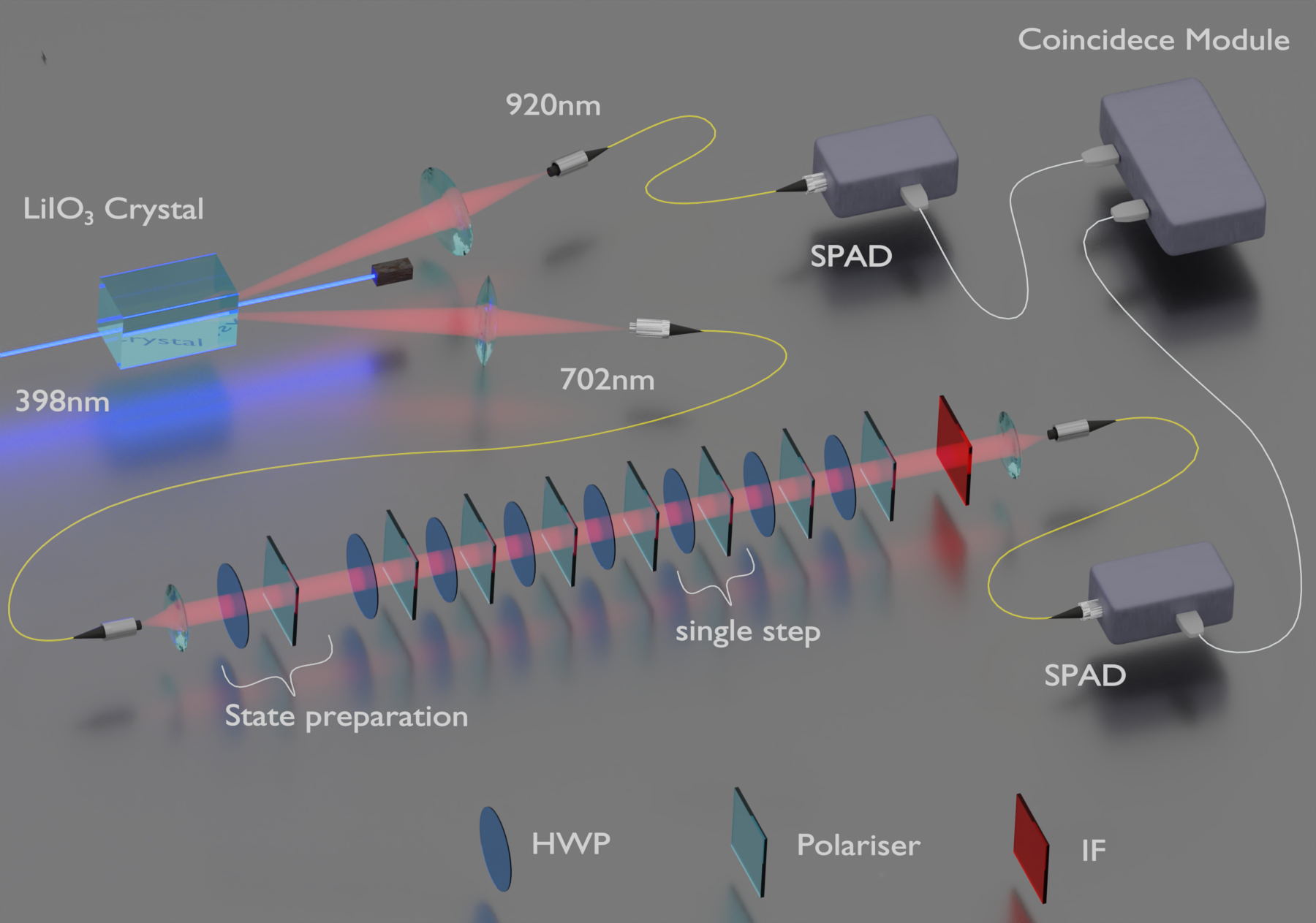}
%\centerline{\epsfig{file=garda7.ps,width=3in}}
\protect\caption{A heralded single-photon source, exploiting type-I parametric downconversion in a LiIO$_3$ crystal, is used to produce single photons at 702 nm.
The photons are collected in a single-mode optical fiber, and then collimated in a 2 mm width Gaussian beam with negligible divergence within the experiment optical path ($\sim2$ m).
A polariser prepares each photon in the horizontal polarisation state.
Then, the photon passes through a series of 7 polarisation rotation/measurement stages.
When measurements are involved (as in the picture), each stage is composed of a half-wave plate (HWP) and a polariser.
In the absence of measurements, a HWP is the only element present in each block (in this case, a polariser is put at the end of the sequence to perform the final projective measurement of the polarisation observable $|H\rangle\langle H|$).
Afterwards, an interference filter (IF, centered at 702 nm, with 3 nm FWHM) is exploited to remove the environmental light; then, the photons are collected by a multi-mode fibre and detected by a silicon single-photon avalanche diode (SPAD).
}
\label{fig2}
\end{figure}

In our experimental implementation (Fig.~\ref{fig2}), heralded single photons (at 702 nm) are generated by a type-I parametric downconversion source.
The photons are collected by a single-mode optical fiber and shaped to form a collimated Gaussian beam (with 2 mm width) over a 2m-long path.
A heralded photon, initialized in the horizontal polarisation state $\ket{H}$, traverses  $N=7$ stages, reproducing the dynamics of the sequence of blocks depicted in Fig.~\ref{fig1}.
Each stage is composed of a half-wave plate (HWP) and a polariser.
%In the $k$-th stage, the HWP is responsible for the combined effect of the rotation sequence $U(\beta_{k-1})$-$U(\Delta\phi_k)$-$U(\alpha_k)$, all belonging to the $k$-th block except for the first one, belonging to the $(k-1)$-th one.
In the $k$-th stage, the HWP is responsible for the combined effect of the global rotation $U(\alpha_{k})U(\Delta\phi_k)U(\beta_{k-1})$.
The polariser, instead, realises the projector $|H\rangle\langle H|$.
These stages, combined in a sequence, reproduce the evolution of a single photon passing through the $N=7$ blocks.
Each of them induces a random photon polarisation rotation $\pm\Delta\phi$ (selected by a random number generator) and realizes the (partially-) selective measurement $\Pi$ in \rqn{2}.
We choose the polarisation rotation jumps such that they have modulus $\Delta\phi = 4^\circ$.
Each jump in the sequence is set to be equal (correlated case) to the one in the previous block with probability $p=\frac{\mathcal{C}+1}{2}$, and opposite (anti-correlated case) with probability $1-p$.
At the end of the sequence, only a horizontally-polarised photon is detected.

We have three possible handles on the photon detection probability: (i) the parameter $\theta$, determining the selective strength of the measurement $\Pi$ in Eq. \rqn{2}; (ii) the correlation coefficient ${\cal C}$); (iii) the amplitude $\Delta\phi$ of the polarisation rotation jumps.
In the experiments we have investigated three cases, $\mathcal{C} = -0.6$, 0 and 0.4, whilst varying $\theta$ and keeping $\Delta\phi$ fixed.
For each value of $\mathcal{C}$, upon initializing the photon in the horizontal polarisation state $\ket{H}$, we have measured the single-photon detection probabilities and standard deviations (uncertainties) of 100 random sequences of $N=7$ polarisation rotation jumps. %, realized by the $N=7$ blocks.
The results show that the photon detection probability decays exponentially with the block index $k$.

\begin{figure}[htb]
\includegraphics[width=0.9\columnwidth]{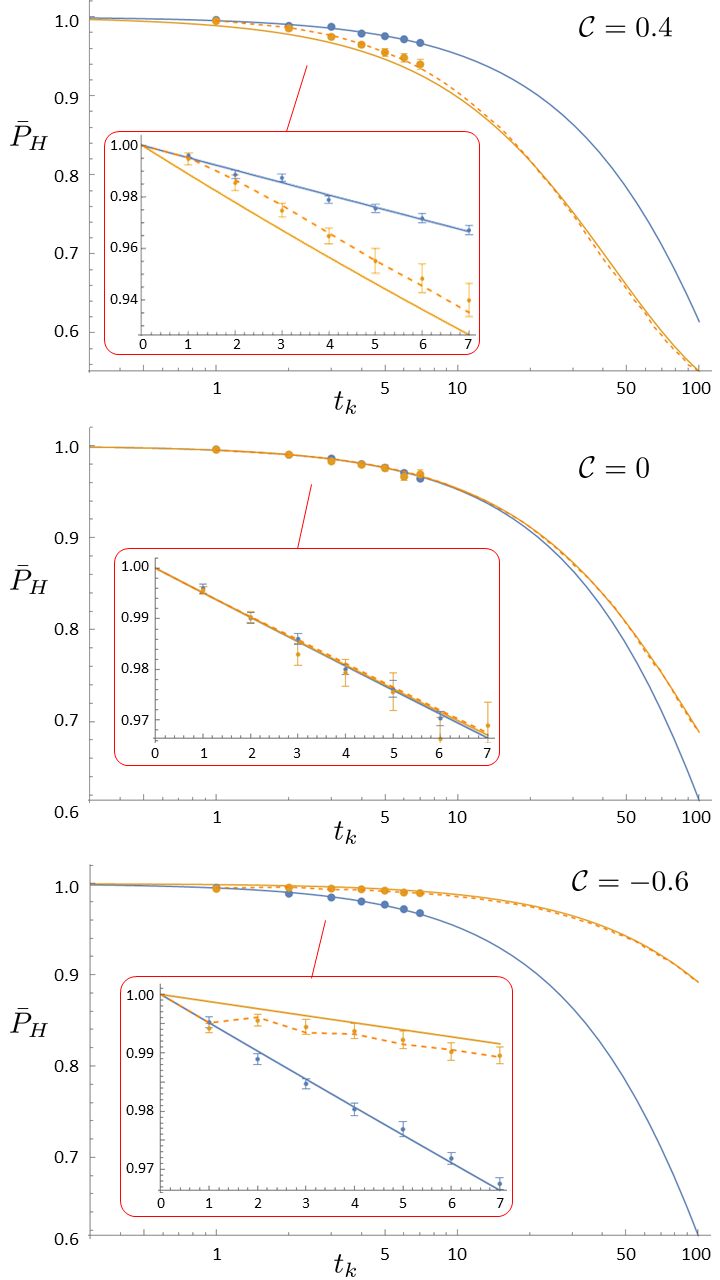}
%\centerline{\epsfig{file=garda7.ps,width=3in}}
\protect\caption{Evolution of the average probability $\bar{P}_H(t_k)$ to detect a single photon with horizontal polarisation for (anti-) correlated consecutive polarisation rotation jumps of magnitude $\Delta\phi=4^\circ$, as a function of $t_k$, for three different values of the correlation parameter $\mathcal{C}$ (top panel: $\mathcal{C} = 0.4$; central panel: $\mathcal{C} = 0$; bottom panel: $\mathcal{C} = -0.6$).
In each plot, the experimental data are reported for both $\theta=1$ (no measurement, in orange) and $\theta=0$ (projective measurement, in blue), with the statistical uncertainties evaluated by taking the standard deviation of 100 different realisations and the solid lines showing the theoretical predictions given by Eqs. \rqn{8.65} and \rqn{8.67}, respectively.
The dashed orange line shows, instead, the numerical estimations for $\bar{P}_H(t_K)$, as per Eq. \rqn{7}, for $\theta=1$, obtained by averaging over 100 simulated random sequences of $\pm\Delta\phi$ jumps.
The inset plots (red boxes) show the detail of the experimentally-investigated region $t_k\in[0,7]$.}
%The solid lines represent the numerical-calculation predictions of the probability $\bar{P}_H$, derived by averaging over 100 different random sequences $\vec{\phi}$, while the dots are the experimentally-measured values of the probability $\bar{P}_H(t_k)$ in correspondence of the discrete time instants $t_k$.
%For each point, the statistical uncertainty is evaluated by taking the standard deviation of the 100 different realisations results.
%The theoretical predictions are given by Eq. \rqn{8.65} for the case $\theta=1$ (no measurement), and by Eq. \rqn{8.67} for the case $\theta=0$ (strong measurement).
\label{fig3}\end{figure}

\begin{figure}[htb]
\includegraphics[width=0.9\columnwidth]{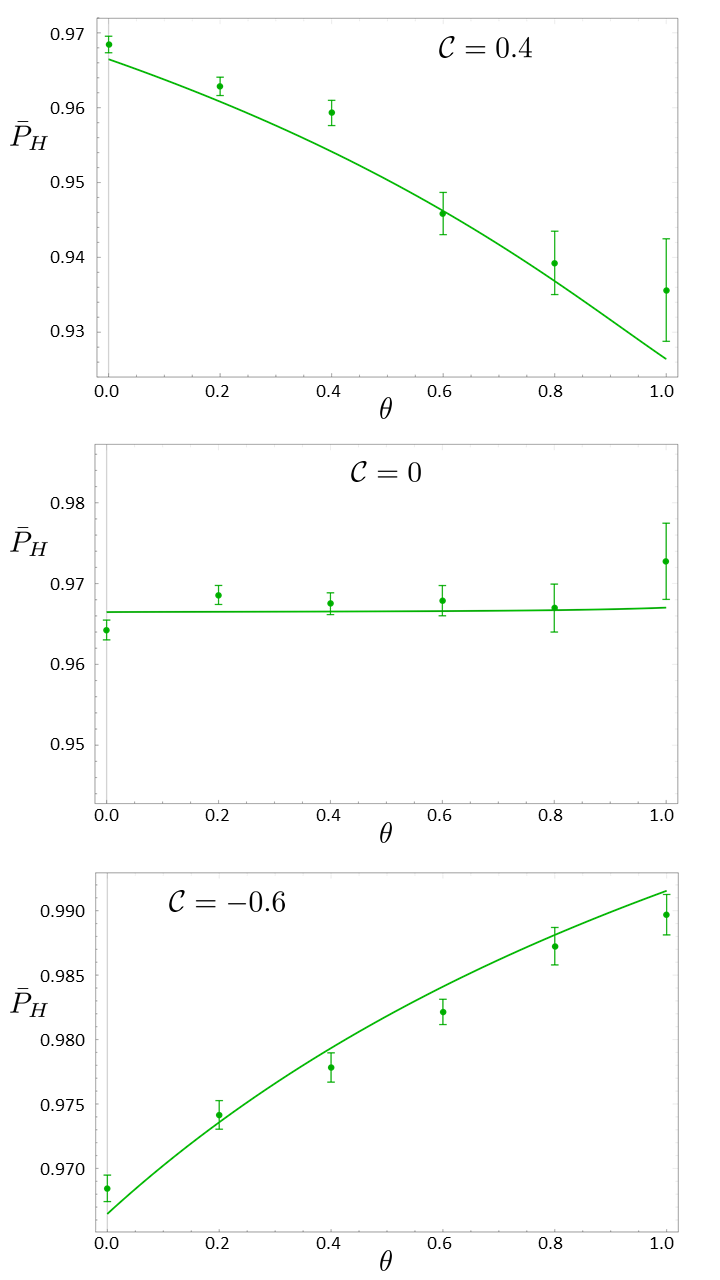}
\protect\caption{Average probability $\bar{P}_H(t_7)$ to detect a single photon with horizontal polarisation $\ket{H}$ at the end of our protocol, as a function of the selective strength $\theta$ of the measurement $\Pi$.
The line represent the theoretical predictions given by Eq. \rqn{8.65'}, by imposing $\tau=1$ and $k=7$, while the dots are the measured values of the probabilities $\bar{P}_H(t_7)$ for different $\theta$ values.
Again, for each experimental point the statistical uncertainty is evaluated by taking the standard deviation of the 100 different realisations results.
}
\label{fig4}\end{figure}

Note that the measurement selective strength decreases as $\theta$ increases.
Thus, Fig.~\ref{fig3} presents the behavior of the photon detection probability with time (i.e., with respect to $t_k$) in two different regimes: $\theta=0$, corresponding to a projective measurement, and $\theta=1$, i.e. in the absence of measurement.
Fig.~\ref{fig4}, instead, shows the results obtained for an intermediate selective strength $0<\theta<1$, i.e. from the projective measurement limit to the weakly-selective measurement one.
In the case $\mathcal{C} = 0$, where the jumps are uncorrelated and the stochastic polarisation angle is a Markovian process, the measured probability is independent of $\theta$.

The novel, hitherto unobserved results are that for $\mathcal{C} > 0\ (\mathcal{C} < 0)$ the QZE (AZE) are revealed, meaning slowdown (speedup) of the decay compared to the uncorrelated case.
By varying $\theta$, one can promote or suppress both effects \cite{21,36,37}.

In both Figs. \ref{fig3} and \ref{fig4}, the obtained results (dots) are in good agreement, within the experimental uncertainties, with the theoretical predictions (lines).\\
%{\bf[In Fig. \ref{fig3} the red lines can be drawn with \rqn{8} and black likes with \rqn{8.49a}.
%In Fig. \ref{fig3}, the lines can be drawn with \rqn{8.65'}, \rqn{8.63}, and \rqn{10}, where $\tau=1$ and $t=7$.]}
%{\bf[FP: we agree, and are already working to modify these figures. This way, we can also remove Fig. \ref{fig10.14}, whose content would be included in Fig. \ref{fig4}.]}

\section{Discussion}
%\textit{Discussion.}
We have experimentally investigated the loss of polarisation coherence of a single photon subjected to a stochastic polarisation noise process (polarisation rotation jumps) and, concurrently, to frequent (weakly or strongly) selective quantum measurements.
We have demonstrated that the polarisation decays at a rate that depends on the non-Markovian correlations within the noise.

More specifically, the key feature revealed by the experiment is the dependence of the decay rate on the correlation parameter: correlated consecutive polarisation rotation jumps give rise to decay slowdown of the decay rate (the QZE), while anti-correlated ones to its speedup (the AZE).
These experimental results fully comply with the KK universal formula for the decay-rate dependence on the overlap between the noise and frequent-measurement spectra \cite{5,22,23,23a,23b,23c}.

This demonstration enables the use of single particles, particularly single photons, as hitherto unexplored probes of noise correlations.
As shown above, a limited number (here, 100) of single-photon polarisation coherence loss events in a noisy medium probed by a few (here, $N=7$) (partially-)selective measurements suffice to reveal unequivocally the noise correlation characteristics.

Characterizing noise, in particular in non-Markovian processes \cite{nmn1,nmn2,nmn3}, is a very important tool for several quantum technologies.
Furthermore, such  probing of the noise may find several fascinating, novel applications.
To this end, the present scheme may be modified by inserting a fluctuating birefringent medium between two consecutive partially-selective measurements \cite{12}, thereby allowing the sensing \cite{q10,q11,q12} of the polarisation noise spectrum associated with a propagating photon.
Among the envisaged applications of the proposed noise sensing method we may contemplate the following:

1)	{\em Probing of magnetic field fluctuations}:
Magnetic field fluctuations can be translated into photon polarisation fluctuations through the Faraday effect \cite{R33,R34,R36,R37}.
The ability to probe such fluctuations by measurements on a single photon may be used to explore the magnetic birefringence of the vacuum \cite{R45}, the parity violation in atoms \cite{R41} and the calibration of weak magnetic noise \cite{R46}.

2) {\em Probing of physiological processes:}
Polarisation microscopy of birefringent cholesterol crystals in human biological (synovial, pleural and pericardial) fluids is an important diagnostic tool of rheumatoid diseases \cite{P0,P1,P3,P4} and atherosclerosis \cite{P5}.
An intriguing, hitherto unexplored question is: can spatial or temporal fluctuations in the concentration of cholesterol crystals in biological fluids and tissues be a novel source of diagnostic information on these diseases?
The ability of single-photon probes to reveal correlations in such fluctuations might be explored by resorting to the results presented here.

To conclude, we have experimentally established the possibility of probing correlations in random fluctuations of photon polarisation by observing QZE and AZE in the polarisation evolution.
Our sensing procedure involves single photons, resulting in a conceptual paradigm shift with respect to interferometric measurements \cite{mwqo,szqo}.

\section{Acknowledgments}
G.K. acknowledges the support of ISF, BSF-NSF, DFG and PACE-IN (QUANTERA).
T.O. acknowledges the support of Czech Science Foundation, grant no. 20-27994S.
This work has received funding from the European Commission’s PATHOS EU H2020 FET-OPEN grant no. 828946 and Horizon 2020.

\appendix*
\section{Dynamical evolution of the photon polarisation}

%In the absence of measurements ($\theta=1$), the average uninterrupted polarisation evolution can be shown \cite{21} to be
%\be
%\bar{P}_H(t_k) =\frac12+\frac{g(r)-g(-r)}{4r},
%\e{8}
%where
%\bea
%&&g(r)=[r+(1-p)\cos(2\Delta\phi)][r+p\cos(2\Delta\phi)]^k,
%\nonumber\\
%&&r=\sqrt{(1-p)^2-p^2\sin^2(2\Delta\phi)}.
%\ea{9}

%We next discuss separately non-random and random evolution under real measurements ($0\le\theta<1$).

%In the case of non-random evolution ($p=1$), Eqs. \rqn{3} and \rqn{4} can be reduced to a closed form for any $\theta$, yielding \cite{21}
%\be
%P_H(t_k)=\frac{(\lambda_1^k(\cos\Delta\phi-\lambda_2)+
%\lambda_2^k(\lambda_1-\cos\Delta\phi))^2}
%{(1+\theta)^2\cos^2\Delta\phi-4\theta},
%\e{11}
%where
%\be
%\lambda_{1,2}=\frac12\left[(1+\theta)\cos\Delta\phi
%\pm\sqrt{(1+\theta)^2\cos^2\Delta\phi-4\theta}\right].
%\e{12}
Let us analyse in more detail the effects of having noisy modulation of the polarisation rotator.
In the case of random evolution ($-1\le\mathcal{C}<1$), for sufficiently small angles $\Delta\phi$, a master equation (ME) can be derived for the evolution of the polarisation probabilities averaged over the realizations of the random process and smoothed over the evolution time \cite{21}.
This ME yields the following rate equations for the average probabilities of the two orthogonal polarisations:
\bea
&&\dot{\bar{P}}_H=-\gamma\bar{P}_H+\gamma\bar{P}_V,\nonumber\\
&&\dot{\bar{P}}_V=\gamma\bar{P}_H-
(\gamma+2\Gamma_0)\bar{P}_V.
\ea{8.62}
Here, $2\Gamma_0$ is the time-averaged photon absorption rate, with
\be
\Gamma_0=-\ln\theta/\tau,
\e{10}
and $\gamma$ denotes the polarisation decay rate:
\be
\gamma=\frac{1+{\cal C}\theta}{1-{\cal {\cal C}}\theta}
\frac{\Delta\phi^2}{\tau}.
\e{8.63}

The solution of Eqs.~\rqn{8.62} for the horizontal component of the photon polarisation is:
\be
\bar{P}_H(t)=e^{-(\gamma+\Gamma_0)t}\left(\cosh(St)+
\frac{\Gamma_0}{S}\sinh(St)\right),
\e{8.65''}
where $S\equiv\sqrt{\gamma^2+\Gamma_0^2}$.
This solution describes the polarisation at the discrete time instants $t_k$.
The validity condition for Eqs. \rqn{8.62}-\rqn{8.65''} is $\gamma\ll\nu$, or \cite{21}
\be
\langle(\Delta\phi_k)^2\rangle\ll(1-{\cal C})(1-{\cal C}\theta),
\e{8.64}
i.e., the polarisation rotation jumps distribution variance should be sufficiently small, and ${\cal C}$ should not be too close to 1.

\subsection*{Polarisation decay rate under random evolution and measurements}

\begin{figure}[htb]
\includegraphics[width=0.9\columnwidth]{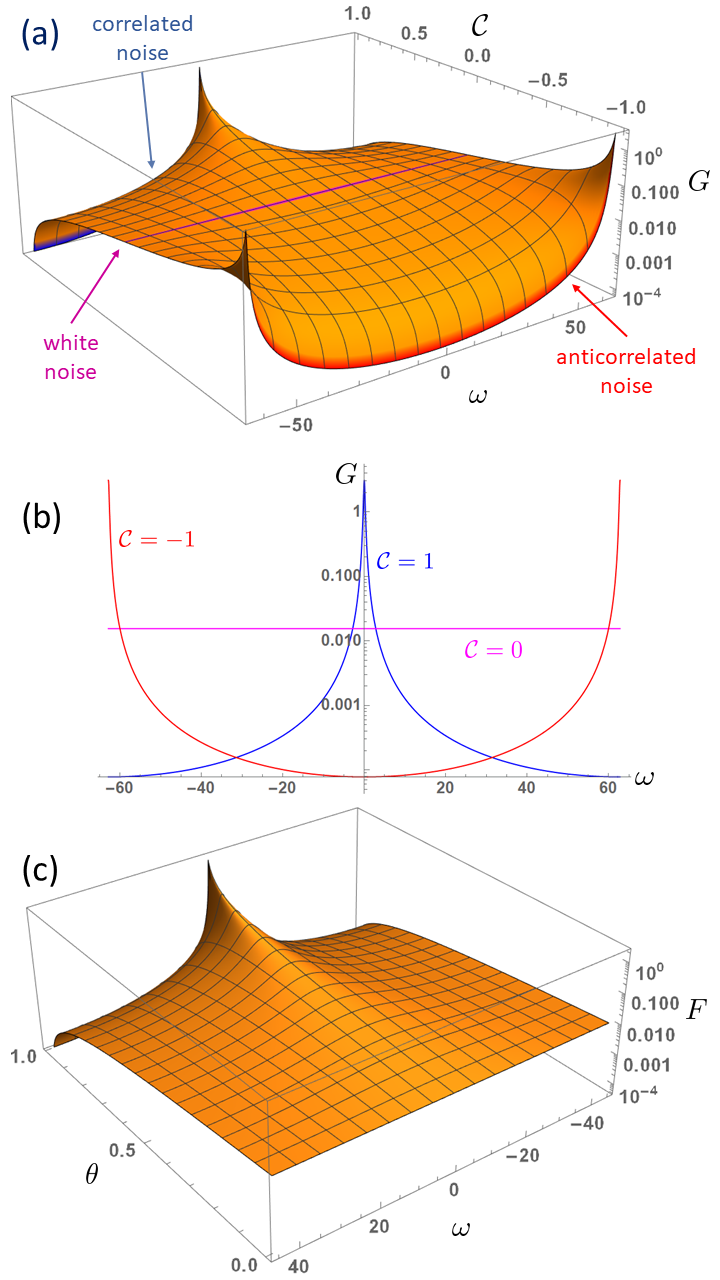}
\protect\caption{Plot (a): spectral density $G(\omega)$ of the rotation-angle fluctuations for $\Delta\phi=4^\circ$ and $\tau=0.05$, as a function of $\omega$ and $\mathcal{C}$, for a time-discrete dynamical evolution of the photon polarisation (see Eq.(\ref{G_omega})).
The maximally correlated ($\mathcal{C}=1$), white ($\mathcal{C}=0$) and anticorrelated ($\mathcal{C}=-1$) noise regions are highlighted in blue, magenta and red, respectively.
Plot (b): detail of the three noise spectral regions highlighted in plot (a), as a function of the noise frequency $\omega$.
Plot (c): measurement control spectrum $F(\omega)$ for $\Delta\phi=4^\circ$ and $\tau=0.05$, as a function of $\omega$ and  $\theta$, again in the time-discrete evolution case (see Eq.(\ref{F_omega})).}
\label{G-F}\end{figure}

The polarisation decay rate obeys the universal formula \rqn{1}, but applied to the discrete-time evolution case identified by the finite integration limits $\pm\pi/\tau$, i.e.:
\be
\gamma=2\pi\int_{-\pi/\tau}^{\pi/\tau}d\omega\,
G(\omega)F(\omega).
\e{8.52}
Here
\be
G(\omega)=\frac{\Delta\phi^2}{2\pi\tau}
\frac{1-{\cal C}^2}{1+{\cal C}^2-
2{\cal C}\cos\omega\tau}
\e{G_omega}
is the spectral density associated with the rotation-angle fluctuations (also called ``bath'' or ``reservoir'' spectrum), whereas the measurement control spectrum
\be
F(\omega)=\frac{\tau}{2\pi}
\frac{1-\theta^2}{1+\theta^2-2\theta\cos\omega\tau}
\e{F_omega}
strictly depends on the transmittance parameter $\theta$.
Once again, it is worth noting that, in the present setup, the frequency domain for $G(\omega)$ and $F(\omega)$ is restricted to the interval $(-\pi/\tau,\pi/\tau)$, since the time evolution is discrete.
Fig.~\ref{G-F}a shows the behavior of $G(\omega)$ as a function of $\omega$ and the correlation coefficient $\mathcal{C}$, with Fig.~\ref{G-F}b illustrating in detail the maximally correlated (in blue), anticorrelated (in red) and white noise (in magenta) cases, while Fig.~\ref{G-F}c shows the dependence of $F(\omega)$ on $\omega$ and the measurement selective strength $\theta$.
In the figure, both $G(\omega)$ and $F(\omega)$ have been plotted for $\Delta\phi=4^\circ$, as in our experimental implementation, and $\tau=0.05$).\\
For {\em highly correlated} jumps $({\cal C}\approx1)$ we find that
\be
G(\omega)\approx\frac{\Delta\phi^2}{\pi\tau^2}
\frac{\Gamma_{\rm B}}{\Gamma_{\rm B}^2+\omega^2},
\e{8.54}
i.e. $G(\omega)$ can be approximated by a narrow Lorentzian of width $\Gamma_{\rm B}=(1-{\cal C})/\tau$.
In Fig.~\ref{G-F}a one can observe how, for {\em partly correlated} random polarisation rotation jumps ($0<{\cal C}<1$), the peak shown by $G(\omega)$ at $\omega=0$ decreases with $\mathcal{C}$, becoming also broader in the process.\\
%\begin{figure}[htb]
%\includegraphics[width=8cm]{fig10-12.jpg}
%%\centerline{\epsfig{file=garda8.ps,width=3.in}}
%\protect\caption{The overlap of $F$ (dashed or dotted lines) and $G$ (solid line) for correlated polarisation-angle (phase) jumps with correlation parameter ${\cal C}=0.7$.
%Dashed line: $F(\omega)$ for $\theta=0$ (perfect projections).
%Dotted line: $F(\omega)$ for $\theta=0.9$ (weak measurements).
%Here $\Delta\phi=0.1,\ \tau=0.07$.}
%\label{fig10.12}\end{figure}
For projective measurements ($\theta=0$), $F(\omega)$ acquires the constant value $\tau/(2\pi)$, whereas for $\theta\ne0$ $F(\omega)$ shows a peaked distribution centered at $\omega=0$, with a characteristic width $\nu=[2\pi F(0)]^{-1}$.
This indicates that the presence of projective measurements causes a {\em reduction} of the $\gamma$ value, which is a signature of the QZE.
However, the QZE signature disappears in the uncorrelated case ${\cal C}=0$, which corresponds to a flat spectral density $G(\omega)=\Delta\phi^2/(2\pi\tau)$.\\
For {\em highly anticorrelated} jumps $({\cal C}\approx -1)$, instead, the bath spectrum $G(\omega)$ is a sum of two shifted narrow Lorentzians of width $\Gamma_{\rm B}'=(1+{\cal C})/\tau$, centered in $\omega=\pm\pi/\tau$, such that:
\be
G(\omega)\approx\sum_{k=\pm 1}\frac{\Delta\phi^2}{\pi\tau^2}
\frac{\Gamma_{\rm B}'}{
{\Gamma_{\rm B}'}^2+(\pi/\tau+k\omega)^2}.
\e{8.55}
In general $G(\omega)$ is peaked at $\omega=\pm\pi/\tau$ for anticorrelated consecutive polarisation-angle jumps ($-1<{\cal C}<0$), see Fig.~\ref{G-F}a.
Such anticorrelated polarisation rotation jumps correspond to the AZE trend, whereby $\gamma$ increases with the measurement rate $\nu$, as opposed to the QZE trend that is observed for correlated jumps.

\bibliographystyle{apsrev4-2}
\bibliography{bibliography.bib}

%\begin{thebibliography}{99}
%\bibitem{}

%\end{thebibliography}

\end{document}